\documentclass{article}

\usepackage{arxiv}

\usepackage[utf8]{inputenc} 
\usepackage[T1]{fontenc}    
\usepackage{hyperref}       
\usepackage{url}            
\usepackage{booktabs}       
\usepackage{amsfonts}       
\usepackage{nicefrac}       
\usepackage{microtype}      
\usepackage{graphicx}
\usepackage[numbers,sort&compress]{natbib}
\usepackage{doi}

\title{Beyond Classification Accuracy: An Exploration-Range Evaluation of Adaptive Crawling for Fake Shopping Sites}

\author{
  Kotono Karasawa\textsuperscript{1} \and
  Kosuke Takeshige\textsuperscript{2} \and
  Shingo Matsugaya\textsuperscript{3,4} \and
  Makoto Shimamura\textsuperscript{3} \and
  Masaki Hashimoto\textsuperscript{1} \\[0.5em]
  \textsuperscript{1}Graduate School of Science for Creative Emergence, Kagawa University \\
  \textsuperscript{2}Chiba Prefectural Police Headquarters \\
  \textsuperscript{3}Trend Micro, Inc. \qquad \textsuperscript{4}Japan Cybercrime Control Center \\[0.3em]
  \texttt{hashimoto.masaki@kagawa-u.ac.jp}
}

\date{}

\renewcommand{\shorttitle}{Exploration-Range Evaluation of Adaptive Crawling}

\hypersetup{
pdftitle={Beyond Classification Accuracy: An Exploration-Range Evaluation of Adaptive Crawling for Fake Shopping Sites},
pdfauthor={Kotono Karasawa, Kosuke Takeshige, Shingo Matsugaya, Makoto Shimamura, Masaki Hashimoto},
pdfkeywords={Fake Shopping Sites, SEO Poisoning, Adaptive Crawling, Closed-Loop Crawler, Seed-Compound Strategy, Exploration-Range Evaluation},
}

\begin{document}
\maketitle

\begin{abstract}
In recent years, fake shopping sites targeting Japanese users have appeared in the top results of search engines through SEO poisoning, causing increasing damage. Conventional collection methods rely on fixed keywords and cannot keep up with evolving attack campaigns, delaying the discovery of new sites. We propose a closed-loop crawler that incorporates the page-level outputs of a fake-site classifier (fastText+LightGBM) into the search queries of the next cycle. Search queries are generated by a seed-compound strategy that combines characteristic words extracted from positive pages with seed words from the fake-shopping context (e.g., ``deep discount,'' ``official''). To complement evaluations that tend to focus on classifier accuracy, we also introduce per-cycle new-host counts and cumulative unique-host counts as exploration-range metrics. In a comparative experiment ($n=3$ for the proposed method, $n=2$ for the baseline), the fixed-keyword baseline yielded zero new-host acquisition from cycle 2 onward, indicating complete stagnation, whereas the proposed method continued to discover new hosts and, at cycle 3, achieved a cumulative unique-host count approximately 7.6 times that of the baseline on average.
\end{abstract}


\section{Introduction}
In recent years, attacks that redirect users from web search results to fraudulent sites such as fake shopping sites have been continuously observed, and public agencies and security vendors have repeatedly issued warnings\cite{jc3shopping,trendmicro_expo,caa_ricefake}. In particular, SEO poisoning is a technique that exploits SEO to make attack-related pages appear in the top results of search engines; it has been reported to redirect users from search queries that combine terms such as ``official,'' ``deep discount,'' and ``in stock'' with timely topics (e.g., Osaka-Kansai Expo 2025, the government rice stockpile)\cite{trendmicro_expo,caa_ricefake}. Since attack campaigns continuously change in both search terms and redirection destinations, one-off investigations and fixed exploration terms tend to delay the discovery of new sites\cite{css2024_Michishita,ACM2014_Leontiadis}.

In this study, we focus on fake shopping sites targeting Japanese users. Among the sites to which users may be redirected through search results under SEO poisoning, we particularly target these fake shopping sites and the stepping-stone pages (compromised legitimate sites) in their redirection chains\cite{css2024_Michishita}. Because these sites can appear in large numbers and have short lifetimes, continuous collection and updating is a prerequisite for analysis and countermeasures. However, conventional collection based on fixed keywords tends to stagnate in finding new sites when search results converge to the same set of sites.

Therefore, the objective of this research is to construct an automatic collection and analysis framework that starts from search results to collect URLs and, by cycling through storage, classification, and exploration-term updates, continues to discover new fake shopping sites while adapting to evolving attacks. The contributions of this work are threefold:
\begin{enumerate}
  \item \textbf{Design of a closed-loop crawler.}
        We design a closed-loop architecture in which the page-level outputs of a classifier (fastText+LightGBM) are fed back into the set of search queries for the next cycle, aiming at fake shopping site discovery. The search queries are generated by a \textit{seed-compound} strategy that combines characteristic words extracted from positive pages with seed words from the fake-shopping context (e.g., ``deep discount'' (\textit{gekiyasu}), ``official'' (\textit{k\=oshiki}), ``in stock'' (\textit{zaiko ari})), maintaining the exploration intent aligned with attack pathways while expanding reach.
  \item \textbf{Introduction of exploration-range evaluation.}
        To complement existing evaluations that tend to focus on classifier accuracy, we introduce per-cycle new-host counts (\texttt{new\_hosts}) and cumulative unique-host counts (\texttt{cum\_unique\_hosts}) as exploration-range metrics. These metrics quantify the ability to ``continue discovering new sites'' under evolving SEO-poisoning attack campaigns.
  \item \textbf{Empirical validation.}
        Through a comparative experiment on Japanese fake shopping sites ($n=3$ for the proposed method, $n=2$ for the baseline), we show that the fixed-keyword approach yields zero \texttt{new\_hosts} from cycle 2 onward, indicating complete stagnation in exploration (cycle 1 also produces only 1.0 hosts on average, which is limited). In contrast, the proposed closed-loop approach continues to acquire new hosts across cycles and, at cycle 3, achieves a cumulative unique-host count approximately 7.6 times that of the fixed-keyword approach on average.
\end{enumerate}

The remainder of this paper is organized as follows. Section~\ref{sec:related} reviews related work and clarifies the position of this study. Section~\ref{sec:design} describes the architecture and processing flow of the closed-loop crawling system, as well as the automatic keyword generation method. Section~\ref{sec:eval} presents the experimental setup and evaluation metrics, and Section~\ref{sec:results} presents the results and an analysis of generated keywords. Section~\ref{sec:discussion} discusses threats to validity, limitations, and operational considerations. Section~\ref{sec:conclusion} concludes the paper and outlines future work.

\section{Related Work}
\label{sec:related}
Related work can be organized into the realities of fake shopping site attacks and their detection (Section~\ref{sec:rel_attack_detection}), the difficulty of observation and collection strategies (Section~\ref{sec:rel_collection}), and the position of this study (Section~\ref{sec:rel_position}).

\subsection{Attack Realities and Detection of Fake Shopping Sites}
\label{sec:rel_attack_detection}
Fake shopping site damages have been continuously reported, with topics and contexts of redirection changing depending on the situation\cite{trendmicro_expo,caa_ricefake,jc3shopping}.
Kodera et al.\ analyzed site structure and operational characteristics\cite{ipsj2021_Kodera}, while Michishita et al.\ reported cases of compromised legitimate sites being exploited as stepping stones from search results to fake shopping sites\cite{css2024_Michishita}; both fake shopping sites (final destinations) and stepping-stone pages (entry points) should therefore be included as targets for collection.
A collection method for fake shopping sites that reuse product information from legitimate sites has also been proposed\cite{css2024_Hasegawa}.

For detection, the field has evolved from learning-based detection using lightweight URL features\cite{ACM2009_Ma} to deep-learning-based URL representation learning\cite{arXiv2018_le}; for Japanese fake shopping sites, a system combining fastText and LightGBM has been proposed\cite{IEEEAccess2023_Sakai}.
Other approaches include visual features\cite{USENIXSecurity2021_Lin}, LLMs with knowledge graphs\cite{arXiv2024_li}, and large-scale fraudulent e-commerce site detection\cite{IEEES&P2023_Bitaab,ACM2017_Claudio}.
As detectors become more sophisticated, the continuity of data collection that underpins evaluation and retraining becomes equally important.

\subsection{Observation Challenges and Collection Strategies}
\label{sec:rel_collection}
Search-engine abuse (SEO poisoning) has long been observed as search-result contamination and redirection attacks\cite{ACM2014_Leontiadis,USENIXAssociation2011_Leontiadis}, where changes in search rankings and redirection destinations can cause stagnation in fixed-keyword collection.
Crawlers also face evasion such as cloaking\cite{ACM2011_Wang,IEEES&P2016_Invernizzi}, and in the phishing field, blacklist evasion measurement\cite{IEEES&P2019_Oest}, large-scale cloaking detection\cite{IEEES&P2021_Zhang}, and responses to CAPTCHA-based serving\cite{USENIXSecurity2024_Teoh} have been reported, suggesting that collection strategies should be evaluated by reach range and stagnation in addition to detection accuracy.

To address these challenges, guided and adaptive collection has been studied\cite{IEEEComputerSociety2012_Invernizzi}, and large-scale detection of defacements associated with search-engine abuse campaigns has also been reported\cite{USENIXSecurity2021_Yang}.
To keep up with adversarial changes, mechanisms that continuously update exploration terms are key, which motivates the closed-loop approach pursued in this study.

\subsection{Position of This Study}
\label{sec:rel_position}
Existing studies on detection models\cite{IEEEAccess2023_Sakai,arXiv2018_le,USENIXSecurity2021_Lin} primarily aim at improving classification accuracy, and the design of collection strategies themselves is not their main concern.
Studies addressing collection strategies\cite{IEEEComputerSociety2012_Invernizzi,USENIXSecurity2021_Yang} target guided collection or large-scale campaign detection, but a closed-loop structure that directly feeds back detection results into next-cycle search-query generation is not necessarily foregrounded.
An existing method for collecting fake shopping sites\cite{css2024_Hasegawa} uses popular product information as search queries but lacks an automatic query-update mechanism based on collection results.
In contrast, this study proposes and evaluates a closed-loop crawler that feeds detection results back into search-query updates, demonstrating effectiveness through a relative comparison via the exploration range (\texttt{cum unique hosts}) and the stagnation metric (\texttt{new hosts}).

\section{System Design}
\label{sec:design}

\subsection{Closed-Loop Architecture}

\begin{figure}[t]
  \centering
  \includegraphics[width=\linewidth]{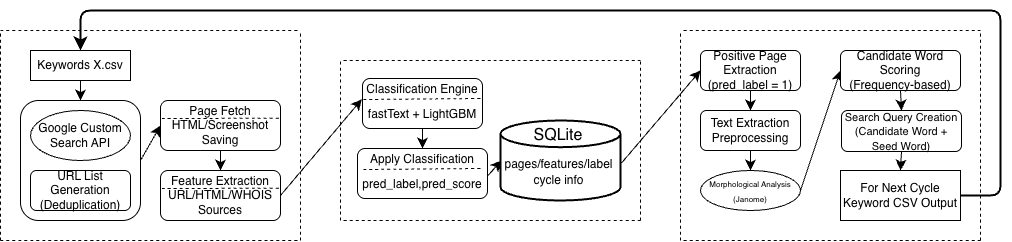}
  \caption{Overall system architecture: a closed-loop of collection $\to$ classification $\to$ search-query update. The page-level outputs of the classifier (fastText+LightGBM) are fed back into the search-query generation of the next cycle.}
  \label{fig:arch}
\end{figure}

The proposed system adopts a closed-loop architecture in which classification results are fed back into the search-query generation of the next cycle. It consists of five modules: (1) a search and collection module, (2) an SQLite database, (3) a feature extraction module, (4) a classification engine (fastText+LightGBM), and (5) an automatic keyword generation module (Figure~\ref{fig:arch}). The feedback of classification results (Module 4 $\to$ 5 $\to$ 1) is the core of this system; below, we describe the design of each module.

Each cycle is executed in the following steps:
\begin{enumerate}
  \item \textbf{Search and collection.} Obtain URLs via the search API and save HTML through HTTP fetching (with browser-based fetching as needed).
  \item \textbf{Classification.} Extract features from the saved HTML, and assign \texttt{pred\_label} (positive/negative) and \texttt{pred\_score} using the trained classifier.
  \item \textbf{Search-query generation (Condition B only).} Extract candidate words from the set of pages with \texttt{pred\_label}=1, and generate a keyword CSV for the next cycle.
  \item \textbf{Storage and aggregation.} Save the database for each cycle and aggregate metrics (unique hosts, etc.).
\end{enumerate}

Collected data are centrally managed in SQLite. The URL, fetch time, cycle number, HTML save path, screenshot save path, \texttt{pred\_label}, and \texttt{pred\_score} are stored, enabling cycle-wise comparison and re-analysis. WHOIS information is unstable in retrieval and prone to missing values; we did not fully leverage it in this experiment (discussed in Section~\ref{sec:whois}).

\subsection{Search and Collection Module}

This module is responsible for sending queries to the search API and fetching the HTML of the resulting URLs. In what follows, we describe the search-result retrieval configuration, URL normalization, and handling of fetch failures and cloaking, in this order.

\paragraph{Search-Result Retrieval Configuration.}
We use the Google Custom Search API\cite{google_api} to retrieve search results. Due to API limits on the number of results per call, the implementation is configured to retrieve 3 results per API call and 2 pages per keyword, so that the theoretical maximum is 6 URLs per keyword (the actual number is less due to deduplication and fetch failures).

\paragraph{URL Normalization and Deduplication.}
The search API may return the same page through different queries, and may also return many URLs that differ only in parameters within the same host. From the perspectives of exploration-range evaluation (at the host level) and collection cost, we therefore designed the system to suppress duplication before and after fetching. Specifically, we introduced (i) URL normalization (scheme unification, trailing-slash unification, removal of known tracking parameters), (ii) URL-level deduplication, and (iii) a per-host upper bound (a cap on the number of fetches from the same host), preventing a single cycle from being disproportionately biased toward the same target. However, overly strong normalization can cause us to miss distinct pages; since this paper's main objective is to compare exploration ranges, we use host novelty (\texttt{new\_hosts}) as the central metric and leave URL-level coverage for future work.

\paragraph{Handling of Fetch Failures and Cloaking.}
Web fetching may yield unstable HTML due to access denial, timeouts, dynamic rendering, or content differences depending on the access source (cloaking). In our implementation, HTTP fetch failures are retried; if they still fail, the reason for failure (HTTP status, exception type) is recorded in the database. We can also switch to a headless browser to save HTML and screenshots, which can be used for visual inspection and downstream analysis. However, since our evaluation is limited to successfully fetched pages, the impact of fetch failures on exploration-range metrics (e.g., bias caused when certain host groups are difficult to fetch) remains. This issue is discussed as a threat to validity in Section~\ref{sec:validity}.

\subsection{Feature Extraction and Classification Engine}
\label{sec:classifier}

For classification, we use URL information, HTML text information, and WHOIS information (when available). Table~\ref{tab:features} shows examples of features. In addition to simple features such as URL length and the ratio of special characters, surface-level features extracted from HTML (e.g., word count) and fastText embeddings are fed into LightGBM.

\begin{table}[t]
\centering
\caption{Examples of features (representative implementation).}
\label{tab:features}
\small
\begin{tabular}{p{0.18\linewidth}p{0.72\linewidth}}
\toprule
Type & Examples (overview) \\
\midrule
URL & URL length, domain length, digit/symbol ratio, TLD type. \\
HTML & Extracted text length, word count, link count (simple), fastText representation, etc. \\
WHOIS & Domain age, registration country, registrar (when available), etc. \\
\bottomrule
\end{tabular}
\end{table}

The classifier follows the framework of prior work: it converts the body text extracted from HTML into a distributed representation (document vector) using fastText, and performs binary classification with LightGBM, combining the URL/HTML additional features\cite{IEEEAccess2023_Sakai}. \textbf{Our claim does not rely on high classifier accuracy.} Since the goal of this research is to evaluate an exploration framework that continues to discover new sites, neither novel proposal of the classifier itself nor its rigorous comparison is within our scope. In the evaluation, we use the same classifier and the same settings for both conditions A and B, and discuss the relative comparison in terms of exploration-range and stagnation metrics.

For reference, the classifier's accuracy on an evaluation set of 42 instances is: Accuracy=0.929, Positive Recall=1.000, Negative Recall=0.625, indicating that false positives may remain (these are reference values for understanding classifier characteristics and do not constitute the basis of this paper's claims). Therefore, our \texttt{pos} counts do not guarantee the true number of fake sites. Nonetheless, since the exploration feedback is generated from ``the positively predicted set,'' classifier errors can affect keyword quality (Section~\ref{sec:kw_quality}).

\subsection{Automatic Keyword Generation}
\label{sec:kwgen}

The automatic keyword generation module takes the set of pages with \texttt{pred\_label}=1 as input, and performs preprocessing (HTML $\to$ text), morphological analysis, candidate-word extraction, scoring, query generation, and CSV output. For morphological analysis, we use Janome\cite{janome}; we adopt nouns primarily and exclude single-character words, symbol-only words, digit-only words, etc.\ to form the candidate-word set. For scoring, to ensure stable operation in small-scale experiments, we use a frequency-based scheme and use the top $K$ words ($K=20$) as the core of the next-cycle exploration.

\paragraph{Candidate-Word Filter.}
The noun sequence extracted by morphological analysis often contains template words and generic e-commerce terms. We therefore introduce a simple rule-based filter that is stable even in small-scale experiments, to suppress the inclusion of overly generic or noisy words. Table~\ref{tab:kwfilter} shows examples of candidate-word filtering. Because filters need updating as exploration targets evolve, we implement the current set as fixed provisional rules in this paper, and discuss improvements in Section~\ref{sec:discussion_kw}.

\begin{table}[t]
\centering
\caption{Examples of candidate-word filtering (representative implementation).}
\label{tab:kwfilter}
\small
\begin{tabular}{p{0.22\linewidth}p{0.68\linewidth}}
\toprule
Item & Example rules \\
\midrule
Length & Exclude words of length 1; symbol-only or digit-only words \\
Part of speech & Primarily nouns (exclude symbols, particles, auxiliary verbs) \\
Normalization & Unify full-width/half-width forms; unify upper/lower case (alphabet) \\
Stop words & Template words (e.g., ``login,'' ``cart'') \\
\bottomrule
\end{tabular}
\end{table}

\paragraph{Seed-Compound Strategy.}
If we use only candidate words extracted and filtered from positive pages as search queries, the inclusion of generic words can cause the search scope to diffuse and drift away from the goal (fake shopping site discovery). To address this, we generate queries that combine a candidate word $w$ derived from a positive page with a seed word $s$ from the fake-shopping context (e.g., ``deep discount,'' ``in stock,'' ``official,'' ``\=odori-yasu,'' ``mail order''). We refer to this as the \textbf{seed-compound strategy}. These seed words were selected based on alerts and investigation reports indicating that fake shopping sites convey trustworthiness and a sense of bargain by emphasizing ``extremely low prices below market rate'' and ``official/authentic'' to encourage purchases\cite{trendmicro_expo,caa_ricefake,jc3shopping}.

\paragraph{Algorithm.}
The generation process is organized as the following algorithm. Let $S_c$ be the set of positive pages at cycle $c$, $T(p)$ the text extracted from page $p$, and $Tok(\cdot)$ the morphological analysis result (noun sequence).
\begin{enumerate}
  \item For each page $p$ in $S_c$, extract $T(p)$, and obtain a candidate-word set $W$ from $Tok(T(p))$.
  \item Apply the rules in Table~\ref{tab:kwfilter} to $W$, and aggregate the occurrence count $f(w)$ of each word.
  \item Select the top $K$ words by $f(w)$ ($K=20$), and combine them with the seed word set $Seed$ to generate queries $Q$ (e.g., $w +$ ``deep discount'').
  \item Output $Q$ as a CSV and use it as the search input for the next cycle.
\end{enumerate}
The frequency-based scheme is easy to implement and works at small scale, but it carries the risks of generic-word contamination and lack of diversity; subsequent improvements (e.g., TF-IDF, phrase extraction, score weighting) are discussed in Section~\ref{sec:discussion_kw}.

\section{Evaluation}
\label{sec:eval}

\subsection{Comparison Conditions}
Condition A (A fixed) explores using six initial keywords fixed throughout cycles 0--3.
The initial keywords (used in cycle 0) were chosen to cover fake e-commerce, tech-support scams, and warning scams: three product-related ones (\texttt{PS5 in stock}, \texttt{iPhone deep discount}, \texttt{Louis Vuitton authentic cheap}), two tech-support ones (\texttt{Apple ID locked}, \texttt{Amazon inquiry}), and one warning one (\texttt{virus warning fake}), totaling six.
Condition B (B boost) uses the same initial keywords as Condition A at cycle 0; from cycle 1 onward, it updates the search using the top 20 keywords automatically generated by the seed-compound strategy described in Section~\ref{sec:kwgen}.
Here, \texttt{boost} refers to the expansion of exploration range (exploration boost) and is unrelated to machine learning methods such as gradient boosting.

Note that the initial keywords also include phrases from attack contexts other than fake shopping (i.e., tech-support and warning scams).
This is intended to confirm a behavior in which, even when exploration begins from diverse entry points, the queries in subsequent cycles converge to the fake-shopping context through the feedback via the classifier (Japanese fake shopping site classification) in the closed loop.
In fact, all generated keywords from cycle 1 onward, shown in Table~\ref{tab:kwexample}, are composed in combination with seed words from the fake-shopping context, confirming the natural narrowing function of the closed loop.

Note also that, since the results of the search API can vary depending on time, both conditions are independently executed multiple times to evaluate them while taking variability into account. The details of independent runs are described in Section~\ref{sec:independent_runs}.

\subsection{Variability Evaluation through Independent Runs}
\label{sec:independent_runs}

Since the results of the search API can vary depending on time and date, in this study we independently ran each condition multiple times as follows. Condition B (B boost) was run three times (B-1, B-2, B-3) and Condition A (A fixed) was run twice (A-1, A-2), independently. Each independent execution (Run) shared the same full configuration---including the initial keyword set, seed-word set, classifier (the trained model of fastText+LightGBM), search-API parameters (region, language, number of results to retrieve, etc.), and the number of cycles (three)---and varied only the \textbf{start date and time of execution}. This setting allows us to evaluate both the reproducibility of the closed loop as a whole under the same configuration and the impact of time-dependent variations on the search API side.

In Section~\ref{sec:results}, we mainly report the cross-Run mean and standard deviation of the main metrics (\texttt{cum\_unique\_hosts}, \texttt{new\_hosts}) for each condition. Per-Run values are reported alongside in the appendix or within tables in the main text.

In this study, Condition B was run three times and Condition A was run twice. Although the sample size is limited, the main claim of this study concerns the comparison of exploration-range metrics between Conditions B and A, and the effect size (B/A ratio) of approximately 7.6 times is so large that the difference between the conditions can be clearly separated even with a limited sample size. Specifically, even in the worst pairwise comparison between individual Runs (B-2/A-1 = 6.7 times), Condition A is exceeded by more than five times, and the best comparison (B-3/A-2 = 8.3 times) shows essentially the same trend (Section~\ref{sec:results}). Therefore, although additional experiments would be required to perform stricter statistical tests (e.g., a t-test), within the experimental scale of this study, we judge that the possibility of the difference between conditions arising by chance is essentially excluded.

\subsection{Collection Pipeline and Settings}
\label{sec:collection_pipeline}
\label{sec:kw_setting}

\paragraph{Pipeline.}
In this study, we define one cycle as ``search $\to$ fetch $\to$ store $\to$ classify $\to$ search-query update.'' Among the processes in each cycle, the factors that directly affect the exploration range (the number of reached hosts) include:
(i) bias in URLs at the top of search results,
(ii) URL duplication,
(iii) fetch failures (denials, timeouts, etc.), and
(iv) collection of multiple pages from the same host.
We therefore apply the following preprocessing policies:

\begin{itemize}
  \item \textbf{URL-level deduplication.} The same URL is stored and classified only once.
  \item \textbf{Host extraction.} The host part is extracted, and exploration-range metrics are aggregated at the host level (\texttt{unique hosts}).
  \item \textbf{Definition of fetch success.} An item is counted as \texttt{total} if its HTML could be saved through HTTP fetching (with browser-based fetching as needed).
  \item \textbf{Storage.} The HTML path, screenshot path, fetch time, cycle number, and classification result are recorded in SQLite.
\end{itemize}

\paragraph{Classifier.}
As described in Section~\ref{sec:classifier}, the classifier follows the framework of prior work\cite{IEEEAccess2023_Sakai}, and \textbf{our claim does not rely on high classifier accuracy}. We use the same classifier and the same settings for both Conditions A and B, and discuss the \textbf{relative comparison in terms of exploration-range and stagnation metrics}. Since our \texttt{pos} is a predicted label rather than the ground truth, classifier errors can affect keyword quality (Section~\ref{sec:kw_quality}).

\paragraph{Keyword generation.}
For Condition B, we apply the seed-compound strategy in Section~\ref{sec:kwgen} with $K=20$, using the set of positive pages (pred label=1) from the previous cycle as input. The seed-word set is the same as in Section~\ref{sec:kwgen}.

\subsection{Evaluation Metrics}
We evaluate the exploration range using the following metrics. Here, the host is extracted from the URL's host part, and uniqueness is determined at the host level.
\begin{itemize}
  \item \texttt{total}: the number of items that were fetched, saved, and became targets of classification (not the total hit count of the search engine).
  \item \texttt{pos}/\texttt{neg}, \texttt{pos rate}: the number and ratio of items predicted as positive (fake candidates) and negative by the classifier.
  \item \texttt{unique hosts}: the number of unique hosts observed in the cycle.
  \item \texttt{new hosts}: the number of new hosts not observed before the current cycle (important for stagnation detection).
  \item \texttt{cum unique hosts}: the cumulative number of unique hosts (the main metric of reach range).
\end{itemize}

To assess the ``efficiency'' of exploration, we also use derived metrics: \texttt{new hosts} / \texttt{total} (new-reach rate) and \texttt{unique hosts} $-$ \texttt{new hosts} (the number of revisited known hosts; degree of duplication).

\section{Results}
\label{sec:results}

\subsection{Per-Cycle Aggregation}
Table~\ref{tab:cycle} shows the per-cycle aggregation results (mean$\pm$SD), and Table~\ref{tab:cycle_runs} shows the per-Run values of \texttt{cum unique hosts}. The aggregation is based on independent runs (Section~\ref{sec:independent_runs}) with $n=2$ for Condition A (A fixed) and $n=3$ for Condition B (B boost).

Condition A (A fixed) observed an average of \texttt{unique hosts}=29.0$\pm$2.8 at cycle 0; from cycle 2 onward, \texttt{new hosts} became 0, with search results converging to the same set of sites and exploration completely stagnating. Consequently, \texttt{cum unique hosts} also remained at 30.0$\pm$1.4 from cycle 2 onward. At cycle 1, a small acquisition of \texttt{new hosts}=1.0$\pm$1.4 was observed, but this is attributable to slight time-dependent fluctuations in the search API results; the trend of early-stage stagnation, with cycle 2 onward becoming completely 0, is common to both runs (Table~\ref{tab:cycle_runs}).

In contrast, Condition B (B boost) continued to reach new regions, with \texttt{new hosts}=85.3$\pm$7.6 at cycle 1 and \texttt{new hosts}=87.3$\pm$5.5 at cycle 2; \texttt{cum unique hosts} reached 229.0$\pm$18.4 at the end of cycle 3. Across all three runs, continuous acquisition of new hosts from cycle 1 to cycle 3 was confirmed (B-1: 237, B-2: 208, B-3: 242), indicating reproducibility under the same configuration. Although \texttt{new hosts} at cycle 3 decelerated to 29.0$\pm$4.4 compared with cycles 1 and 2, continuous discovery of new hosts was nevertheless maintained.

In terms of the mean comparison of \texttt{cum unique hosts}, Condition B reached approximately \textbf{7.6 times} that of Condition A (229.0/30.0). This indicates that updating search queries based on classifier outputs can avoid the stagnation (\texttt{new hosts}=0) that tends to occur in search-based collection and incrementally expand the range of reachable hosts, as demonstrated reproducibly across multiple independent runs.

Furthermore, while \texttt{pos rate} stayed around 0.20--0.23 for Condition A, Condition B maintained higher levels: 0.34$\pm$0.04 at cycle 1, 0.38$\pm$0.08 at cycle 2, and 0.33$\pm$0.05 at cycle 3. Although our \texttt{pos} is a predicted label rather than the ground truth, as a relative comparison under the same classifier and the same settings, this suggests that search-query updates not only expand the reach but also reach regions where positive predictions are more likely to be obtained. Note that \texttt{pos rate} showed some variability across runs (especially pronounced in Condition A), confirming that even under the same configuration, time-dependent fluctuations in the search API can affect the proportion of pages predicted as positive.

\begin{table}[t]
\centering
\caption{Per-cycle aggregation results for each condition (mean$\pm$SD; Condition A: $n=2$, Condition B: $n=3$).}
\label{tab:cycle}
\small
\begin{tabular}{lrrrrrr}
\toprule
Condition & cycle & total & pos rate & unique hosts & new hosts & cum unique hosts \\
\midrule
A fixed ($n$=2) & 0 & 35.5$\pm$0.7  & 0.20$\pm$0.12 & 29.0$\pm$2.8 & 29.0$\pm$2.8 & 29.0$\pm$2.8 \\
A fixed ($n$=2) & 1 & 35.5$\pm$0.7  & 0.23$\pm$0.12 & 29.0$\pm$2.8 & 1.0$\pm$1.4  & 30.0$\pm$1.4 \\
A fixed ($n$=2) & 2 & 35.5$\pm$0.7  & 0.21$\pm$0.10 & 29.0$\pm$2.8 & 0$\pm$0      & 30.0$\pm$1.4 \\
A fixed ($n$=2) & 3 & 35.5$\pm$0.7  & 0.23$\pm$0.12 & 29.5$\pm$2.1 & 0$\pm$0      & 30.0$\pm$1.4 \\
\midrule
B boost ($n$=3) & 0 & 35.0$\pm$1.7  & 0.22$\pm$0.10 & 27.3$\pm$3.2 & 27.3$\pm$3.2 & 27.3$\pm$3.2  \\
B boost ($n$=3) & 1 & 109.7$\pm$3.8 & 0.34$\pm$0.04 & 90.3$\pm$6.0 & 85.3$\pm$7.6 & 112.7$\pm$9.2 \\
B boost ($n$=3) & 2 & 115.0$\pm$3.6 & 0.38$\pm$0.08 & 98.0$\pm$6.1 & 87.3$\pm$5.5 & 200.0$\pm$14.2 \\
B boost ($n$=3) & 3 & 39.3$\pm$2.9  & 0.33$\pm$0.05 & 37.3$\pm$1.2 & 29.0$\pm$4.4 & 229.0$\pm$18.4 \\
\bottomrule
\end{tabular}
\end{table}

\begin{table}[t]
\centering
\caption{Per-Run values of \texttt{cum unique hosts}.}
\label{tab:cycle_runs}
\small
\begin{tabular}{lrrrr}
\toprule
Run & cycle 0 & cycle 1 & cycle 2 & cycle 3 \\
\midrule
A-1 & 31 & 31  & 31  & 31  \\
A-2 & 27 & 29  & 29  & 29  \\
\midrule
B-1 & 31 & 118 & 205 & 237 \\
B-2 & 25 & 102 & 184 & 208 \\
B-3 & 26 & 118 & 211 & 242 \\
\bottomrule
\end{tabular}
\end{table}

\subsection{Derived Metrics}
\label{sec:derived_metrics_results}
Table~\ref{tab:derived} summarizes the derived metrics defined in Section~\ref{sec:eval}: new-reach rate (\texttt{new hosts} / \texttt{total}) and known revisits (\texttt{unique hosts} $-$ \texttt{new hosts}).

For Condition A, from cycle 2 onward, the new-reach rate is 0 and known revisits remain at about 29, indicating that the search results are completely fixed within the known set. The new-reach rate at cycle 1 is 0.03$\pm$0.04, which is extremely low, indicating that stagnation is established early. In contrast, for Condition B, the new-reach rate remains at high levels of 0.74--0.78 throughout cycles 1--3, demonstrating that the search-query updates strongly contribute to new-host reaching. Known revisits transition to 10.7$\pm$3.2 at cycle 2 and 8.3$\pm$5.1 at cycle 3, with a particular increase at cycle 2. This suggests that some keywords may have returned to known clusters, which is related to keyword quality (genericization and templatization); this is discussed in Section~\ref{sec:kw_quality}.

\begin{table}[t]
\centering
\caption{Derived metrics (new-reach rate and number of known revisits, mean$\pm$SD).}
\label{tab:derived}
\small
\begin{tabular}{lrrr}
\toprule
Condition & cycle & new hosts / total & Known revisits (unique - new) \\
\midrule
A fixed ($n$=2) & 0 & 0.82$\pm$0.06 & 0$\pm$0       \\
A fixed ($n$=2) & 1 & 0.03$\pm$0.04 & 28.0$\pm$4.2  \\
A fixed ($n$=2) & 2 & 0$\pm$0       & 29.0$\pm$2.8  \\
A fixed ($n$=2) & 3 & 0$\pm$0       & 29.5$\pm$2.1  \\
\midrule
B boost ($n$=3) & 0 & 0.78$\pm$0.07 & 0$\pm$0       \\
B boost ($n$=3) & 1 & 0.78$\pm$0.06 & 5.0$\pm$1.7   \\
B boost ($n$=3) & 2 & 0.76$\pm$0.06 & 10.7$\pm$3.2  \\
B boost ($n$=3) & 3 & 0.74$\pm$0.15 & 8.3$\pm$5.1   \\
\bottomrule
\end{tabular}
\end{table}

\subsection{Generated Keywords across Cycles}
Table~\ref{tab:kwexample} shows examples of keywords generated under Condition B (the top 3 actually used for search in each cycle, for all 3 Runs). The lexical composition varied substantially across Runs, reflecting time-dependent variations in search-API results that cause the initial positively predicted page set to differ across Runs. Nonetheless, across all Runs, the composite structure of ``candidate word + seed word'' from the fake-shopping context (``deep discount,'' ``in stock,'' ``authentic,'' etc.) was consistently preserved, demonstrating that the seed-compound strategy functions robustly against variations in initial conditions and underlies the stable exploration-range expansion observed in Table~\ref{tab:cycle}.

\begin{table}[t]
\centering
\caption{Examples of generated keywords under Condition B (top 3 per cycle, 3 Runs).}
\label{tab:kwexample}
\small
\begin{tabular}{lp{0.25\linewidth}p{0.25\linewidth}p{0.25\linewidth}}
\toprule
cycle & B-1 & B-2 & B-3 \\
\midrule
1 & Casino deep discount / Casino in stock / Casino authentic & iPhone deep discount / iPhone in stock / iPhone authentic & shipping deep discount / shipping in stock / shipping authentic \\
2 & favorites deep discount / shipping deep discount / shipping in stock & ``-iri'' deep discount / size deep discount / shipping deep discount & case deep discount / order deep discount / top deep discount \\
3 & top deep discount / supplies deep discount / top in stock & case deep discount / iPhone deep discount / ballpoint pen deep discount & printing deep discount / signboard deep discount / fee deep discount \\
\bottomrule
\end{tabular}
\end{table}

On the other hand, across all Runs, generic e-commerce terms and common words such as ``favorites,'' ``shipping fee,'' ``top,'' ``case,'' and ``order'' appeared in the top positions. While such words contribute to expanding the exploration range, even when combined with seed words they do not necessarily improve redirection efficiency to fake sites, and tend to return to known clusters. The fact that, according to the derived metrics (Table~\ref{tab:derived}), known revisits show an increasing trend from cycle 2 onward (cycle 2: 10.7$\pm$3.2, cycle 3: 8.3$\pm$5.1) is consistent with this. Therefore, jointly achieving exploration-range expansion and maintenance of the positive rate requires scoring and filter design that suppresses the generality of the generated words.

\subsection{Summary}
From the above, the fixed-keyword approach (Condition A) stagnates early due to the convergence of search results: \texttt{cum unique hosts} remained at 30.0$\pm$1.4 at cycle 3, as confirmed across 2 Runs. In contrast, the keyword-expansion approach (Condition B) continuously acquired \texttt{new hosts} across multiple cycles, and expanded \texttt{cum unique hosts} to 229.0$\pm$18.4 at cycle 3 (confirmed across 3 Runs). This corresponds to approximately \textbf{7.6 times} the value of the fixed-keyword approach, demonstrating that search-query updates contribute strongly to exploration-range expansion, reproducibly across multiple independent runs. Furthermore, since \texttt{pos rate} remained at relatively high levels, search-query updates may also contribute to reaching regions that include positively predicted pages.

To confirm the robustness of this value of approximately 7.6 times, we compared \texttt{cum unique hosts} across the individual Runs of Condition B's 3 Runs and Condition A's 2 Runs (Table~\ref{tab:run_ratio}). Across all six pairwise comparisons, the ratio ranged from 6.7 times (B-2/A-1) at the minimum to 8.3 times (B-3/A-2) at the maximum, falling within the range of \textbf{6.7--8.3 times}. Therefore, the effect of the proposed method exceeding the fixed-keyword approach is stably observed regardless of which individual Runs are chosen.

\begin{table}[t]
\centering
\caption{Pairwise ratios of \texttt{cum unique hosts} between individual Runs (cycle 3, all six pairs).}
\label{tab:run_ratio}
\small
\begin{tabular}{lccc}
\toprule
 & B-1 (=237) & B-2 (=208) & B-3 (=242) \\
\midrule
A-1 (=31) & 7.6 & 6.7 & 7.8 \\
A-2 (=29) & 8.2 & 7.2 & 8.3 \\
\bottomrule
\end{tabular}
\end{table}

However, both \texttt{total} and \texttt{pos rate} varied across cycles and across Runs, suggesting that they may be affected by multiple factors such as search-ranking changes, query genericization, URL duplication and fetch failures, and API constraints. Therefore, when evaluating an exploration strategy, it is important to base the comparison not on \texttt{total} alone, but primarily on stagnation metrics (\texttt{new hosts}) and reach-range metrics (\texttt{cum unique hosts}).

\section{Discussion}
\label{sec:discussion}

\subsection{Stagnation Avoidance and Exploration-Range Expansion}
\label{sec:expansion}
The fact that \texttt{new hosts} became 0 from cycle 2 onward in Condition A suggests a situation in which the search results repeatedly converged to the same set of sites, making it difficult to expand into a new exploration space. The \texttt{new hosts} at cycle 1 is also limited at 1.0$\pm$1.4 on average, indicating that stagnation is established early. While search-based collection is powerful, when the top of the search rankings is fixed to the same cluster, the exploration quickly reaches a plateau\cite{ACM2014_Leontiadis,css2024_Michishita}.

In contrast, in Condition B, search-query updates based on classification results enabled \texttt{new hosts} to continue, and \texttt{cum unique hosts} increased to an average of 229.0$\pm$18.4. Continuous acquisition of new hosts from cycle 1 to cycle 3 was observed across all three Runs (Table~\ref{tab:cycle_runs}), and the effect of exploration-range expansion is reproducibly established across independent runs under the same configuration. Since the new-reach rate in Table~\ref{tab:derived} is high (0.74--0.78 in cycles 1--3), it can be interpreted that the search-query updates worked not by increasing the number of collected pages, but by raising the probability of reaching new hosts.

However, at cycle 3, \texttt{new hosts} decelerated to 29.0$\pm$4.4 on average, clearly slower than cycle 1 (85.3$\pm$7.6) and cycle 2 (87.3$\pm$5.5), and this trend was common across all three Runs. This indicates that the closed-loop exploration expansion does not continue indefinitely; in the current implementation of the seed-compound strategy, due to the limits of keyword quality (mixing of generic words), exploration may gradually return to known clusters (discussed in detail in Section~\ref{sec:kw_quality}). Therefore, the exploration-range expansion effect of this implementation is clearly established at least at the cycle-3 stage, while keyword quality improvement is a challenge for longer-term continuity.

\subsection{Factors of Total Variation and Operational Implications}
\label{sec:total_variation}
In Condition B, \texttt{total} varied across cycles. In particular, at cycle 3, \texttt{total} was clearly smaller than at other cycles across all three Runs (cycle 1: 109.7$\pm$3.8, cycle 2: 115.0$\pm$3.6, vs. cycle 3: 39.3$\pm$2.9), and it is likely that practical operational factors such as constraints on the number of usable search keywords and API constraints overlapped.

The fact that \texttt{total} at cycle 3 dropped to about 1/3 of cycles 1 and 2 across all three Runs suggests that this is not an accidental operational constraint of a single Run, but that structural factors inherent to the closed loop itself may also contribute. Specifically, conceivable factors include a decrease in the number of effective candidate words extracted from the positive page set of cycle 2 (which serves as the generation source of cycle 3 search queries), an increase in the duplication of candidate words, and the return to known clusters causing the diversity of the keyword set to decrease. This corresponds to the room for keyword quality improvement described in Section~\ref{sec:kw_quality}. However, within the experimental scale of this study, it is difficult to quantitatively separate the structural and operational factors, and disentangling them in larger-scale experiments is a challenge for future work.

The deceleration tendency of \texttt{new hosts} also shows similar structurality. In Condition B, the transition of \texttt{new hosts} from cycle 1 $\to$ cycle 2 $\to$ cycle 3 is 85.3 $\to$ 87.3 $\to$ 29.0 on average, maintaining a high level up to cycle 2 and decreasing to about 1/3 at cycle 3 (the cycle 3 / cycle 1 ratio is approximately 0.34). This transition pattern is common across all three Runs (B-1: 87 $\to$ 87 $\to$ 32, B-2: 77 $\to$ 82 $\to$ 24, B-3: 92 $\to$ 93 $\to$ 31), indicating that the inter-cycle attenuation has structural causes attributable to the keyword generation process on the implementation side. Specifically, in the initial cycles (cycles 1 and 2) of the closed loop, diverse candidate words are extracted from diverse positive page sets, whereas in subsequent cycles, it becomes more difficult to extract new candidate words from the existing positive host set; we refer to this as ``local saturation of the exploration space.'' Therefore, while the current closed-loop implementation shows a clear expansion effect at least within three cycles, for longer-term continuity, mechanisms that suppress the depletion of new candidate words (e.g., the contrast-set introduction and diversity constraints in Section~\ref{sec:kw_quality}) become important.

In inter-cycle comparisons, instead of using \texttt{total} alone, it is necessary to interpret the transitions of \texttt{new hosts} and \texttt{cum unique hosts} as primary indicators. This is because, in addition to search APIs only being able to retrieve a portion of the top results, the following factors overlap: (i) URL duplication across multiple keywords, (ii) fetch failures such as access denials and timeouts, (iii) URL-level deduplication, and (iv) API rate limits\cite{google_api}. In exploration-range evaluation, \texttt{new hosts} / \texttt{cum unique hosts} should be emphasized over \texttt{total} alone. Furthermore, on the operational side, designs that balance cost and observability, such as retrying fetch failures and selecting browser-based fetching (rendering only suspicious pages), are important\cite{IEEES&P2019_Oest,IEEES&P2021_Zhang}.

\subsection{Improving Keyword Quality}
\label{sec:discussion_kw}
\label{sec:kw_quality}
While the frequency-based scheme in Section~\ref{sec:kw_setting} is easy to implement and works at small scale, it tends to bring template words and generic e-commerce terms to the top.
In this situation, the search terms become generic; while the diversity of search results may increase, the positive rate may decrease.
In our experiments, the \texttt{pos rate} remained at a stable level, but known revisits (\texttt{unique} $-$ \texttt{new}) show an increasing trend across all three Runs from cycle 2 onward (cycle 2: 10.7$\pm$3.2, cycle 3: 8.3$\pm$5.1, Table~\ref{tab:derived}), suggesting that some queries may have returned to known clusters. This is consistent with the structural factors of \texttt{new hosts} deceleration at cycle 3 described in Section~\ref{sec:expansion} and the \texttt{total} decrease described in Section~\ref{sec:total_variation}, supporting the need for keyword quality improvement.

Promising improvements include introducing a contrast set using pred label=0 to compute TF-IDF or log-likelihood ratio for prioritizing words characteristic of the positive side, phrase formation via noun concatenations (2-gram/3-gram) to capture specific named entities rather than generic terms, score weighting based on pred score to prioritize words from high-confidence pages, and diversity constraints (e.g., MMR) to avoid similar words clustering at the top. These are expected to contribute to jointly achieving exploration-range expansion and maintenance of the positive rate.

\subsection{Threats to Validity and Limitations}
\label{sec:validity}
\label{sec:whois}

\paragraph{Validity threats.} Since this experiment relies on the search API, it can be affected by search-ranking changes, region/language settings, and personalization. For the evaluation of inter-Run variability, we addressed this by independently running Condition B for 3 Runs and Condition A for 2 Runs under the same configuration with fixed search settings (region, language, period; UA, rendering), and reporting the main metrics as mean$\pm$SD (Section~\ref{sec:independent_runs}). As a result, while \texttt{cum unique hosts} shows some variability at 229.0$\pm$18.4 in Condition B, the difference from Condition A (30.0$\pm$1.4) is clearly maintained, and the main message (exploration-range expansion by the proposed method) is reproduced across multiple trials.

Since \texttt{pos} is a predicted label rather than the ground truth, the discussion of exploration expansion is positioned as ``a relative comparison under the same classifier and the same settings.'' Cross-checking with the ground truth (sample precision evaluation through manual verification) is left for future work.

We provide an additional remark on sample size. Condition B with 3 Runs and Condition A with 2 Runs are limited, and stricter statistical verification through larger-scale independent runs is a challenge for future work. However, the main claim of this study, namely ``exploration-range expansion through search-query updates,'' is based on a large effect size of approximately 7.6 times; even in the worst-case pairwise comparison between individual Runs (B-2/A-1 = 6.7 times), Condition A is still substantially exceeded. Therefore, within the experimental scale of this study, the limitation of sample size is unlikely to overturn the conclusion.

\paragraph{Cloaking and dynamic pages.} For stepping-stone sites and fake e-commerce sites, content can change depending on access conditions through cloaking, and the actual nature of pages can be hard to read from static HTML due to dynamic generation\cite{css2024_Michishita,IEEES&P2021_Zhang,IEEES&P2016_Invernizzi}.
In this study, we perform browser-based fetching and screenshot saving as needed, but full browser rendering is costly.
Future challenges include (i) detecting differences between HTTP fetch results and rendering results, (ii) detecting signals such as CAPTCHA, and (iii) interactive fetching only for suspicious pages, to improve observability\cite{USENIXSecurity2024_Teoh}.

\paragraph{WHOIS information.} In our experiments, WHOIS retrieval frequently failed, and analysis based on WHOIS information could not be sufficiently performed. WHOIS information may be useful for tasks such as extracting short-lived domains and characterizing biases in registration country and registrar; we leave for future work improving the retrieval success rate (through retry, caching, and TLD-specific handling) and designing features and metrics that assume missing values.

\section{Conclusion}
\label{sec:conclusion}
In this work, we proposed and evaluated a closed-loop crawling framework for fake shopping site discovery, in which the page-level outputs of a classifier (fastText+LightGBM) are fed back into the search queries of the next cycle through a seed-compound strategy, and introduced per-cycle new-host counts and cumulative unique-host counts as exploration-range metrics. In a comparative experiment ($n=3$ for the proposed method, $n=2$ for the baseline), the fixed-keyword approach yielded zero new-host acquisition from cycle 2 onward, while the proposed method achieved a cumulative unique-host count approximately 7.6 times that of the baseline at cycle 3, reproducibly observed across multiple independent runs.

The multi-run evaluation revealed two new insights: the seed-compound strategy maintained the ``candidate word + seed word'' structure across runs with stable exploration-range expansion of 6.7--8.3 times, demonstrating robustness against initial-condition variations; and a common deceleration of \texttt{new\_hosts} at cycle 3, which we characterize as ``local saturation of the exploration space,'' indicates that closed-loop expansion is limited to approximately three cycles in the current implementation. The contributions lie along two theoretical axes: the closed-loop architecture, which complements detection-model-focused studies by feeding detection results back into next-cycle query generation, and the exploration-range evaluation, which complements accuracy-based assessments by quantifying ``the ability to continue discovering new sites.''

Remaining challenges include validating the classifier through sample-precision evaluation and robustness to distribution shift; improving keyword generation quality via TF-IDF or log-likelihood ratio, n-gram phrase generation, and MMR-based diversity assurance to address ``local saturation''; improving robustness against cloaking and dynamic pages; and stable WHOIS retrieval for short-lived domain extraction and registrar/country bias characterization.

\section*{Acknowledgments}
We acknowledge the use of Claude AI assistant (Anthropic) for the English translation of this paper.

\bibliographystyle{unsrtnat}
\bibliography{refs}

\end{document}